\documentclass[11pt]{article}
\usepackage[a4paper]{geometry}

\usepackage{amsmath, amsthm, amssymb}
\usepackage{dsfont}
\usepackage{graphicx}
\usepackage{subcaption}
\usepackage[round]{natbib}
\usepackage{url}
\usepackage[hidelinks,colorlinks=true,linkcolor=blue,citecolor=blue,urlcolor=blue]{hyperref} 
\usepackage[ruled]{algorithm2e}
\usepackage{booktabs}
\usepackage{placeins} 

\newcommand{\grf}{\texttt{grf~}}

\newcommand{\R}{\texttt{R~}}

\newcommand{\one}[1]{\ensuremath{\mathds{1}\!\left\{#1\right\}}}

\newcommand{\PP}[2][]{\mathbb{P}_{#1}\left[#2\right]}

\author{
  Erik Sverdrup\textsuperscript{1} \and
  James Yang\textsuperscript{2} \and
  Michael LeBlanc\textsuperscript{3,4} \and\\
  \small \textsuperscript{1}Department of Econometrics \& Business Statistics, Monash University\\
  \small \textsuperscript{2}Department of Statistics, Stanford University\\
  \small \textsuperscript{3}Public Health Sciences, Fred Hutchinson Cancer Center\\
  \small \textsuperscript{4}Department of Biostatistics, University of Washington
}

\date{}
\title{Efficient Log-Rank Updates for Random Survival Forests}

\begin{document}
\maketitle

\begin{abstract}
Random survival forests are widely used for estimating covariate-conditional survival functions under right-censoring. Their standard log-rank splitting criterion is typically recomputed at each candidate split. This $O(M)$ cost per split, with $M$ the number of distinct event times in a node, creates a bottleneck for large cohort datasets with long follow-up. We revisit approximations proposed by \citet{leblanc1995step} and develop simple constant-time updates for the log-rank criterion. The method is implemented in \texttt{grf} for \texttt{R} and reduces training time on large datasets while preserving predictive accuracy.
\end{abstract}

\section{Introduction}
Random forests \citep{breiman2001random} are a staple of the machine learning toolbox, often excelling at conditional mean prediction tasks on tabular data. Since their introduction, Breiman's algorithmic blueprint has been extended beyond regression to a wide range of statistical quantities. For survival analysis, \citet{ishwaranRSF} introduced random survival forests to estimate the conditional survival function $S(t; x) = \PP{T_i > t \mid X_i = x}$.
A random survival forest typically proceeds in two phases. First, the covariate space is recursively partitioned via axis-aligned cuts, with splits chosen to maximize a criterion that separates units into groups with differing survival. Second, the resulting partitions are used to produce covariate-specific predictions, either by ensemble-aggregating or kernel-weighting Kaplan--Meier (or Nelson--Aalen) survival functions. The most common splitting criterion is the log-rank statistic, initially suggested for survival trees by \citet{leblanc1993survival, segal1988regression}. Intuitively, the log-rank statistic assesses whether a candidate split separates samples into regions with different survival distributions.

A major reason for the computational success of Breiman's regression forests is the efficiency of CART splitting, which reduces to computing running means \citep{hastie2009elements}. Given sorted samples, evaluating all $n$ candidate split points takes $O(n)$ time. Similar efficiency gains underlie many practical extensions of forests, such as the use of gradient-based approximations in boosted trees \citep{friedman2001greedy} or the generalized random forest framework of \citet{athey2019generalized}, where custom targets defined via estimating equations are reduced to CART splits on gradient-based pseudo outcomes. In contrast, random survival forests do not typically admit such simplifications: the log-rank statistic requires summing over the $M$ distinct event times in a node, giving $O(M)$ computation for every candidate split. To the best of our knowledge, leading implementations, including \texttt{randomForestSRC} \citep{ishwaranRSF}, \texttt{ranger} \citep{wright2015ranger}, and \texttt{sksurv} \citep{polsterl2020scikit}, all implement log-rank splitting in this way, as do other approaches relying on two-sample tests for split evaluation \citep{hothorn2003exact}.

These computational demands are especially challenging for large datasets with long follow-up, common in modern cohort studies.
For example, the SEER cancer database \citep{SEER} follows patients for several decades and records numerous high-cardinality predictors, such as continuous measures of mitotic rate.
In these settings, modern statistical learning methods are increasingly deployed to estimate nuisance components for causal quantities \citep[see, e.g.,][]{wagerbook}.
For instance, when estimating heterogeneous treatment effects or regression discontinuity parameters in time-to-event data with right-censoring (e.g., \citet{cui2023estimating, schuessler2026rdd}), accounting for censoring is necessary for unbiased estimation.
When flexible machine learning estimators, such as random survival forests, are used, this is typically achieved via doubly robust censoring adjustments \citep{robins1994estimation, rubin2007doubly}. 
These adjustments require estimating two conditional survival functions: one for the event process and one for the censoring process.

To mitigate computational burden, current practice often relies on heuristics, such as restricting the number of events considered, subsampling split points, or screening covariates. Motivated by this bottleneck, we seek an algorithmic solution closer in spirit to Breiman's original one-pass scanning step by leveraging approximations.
Deriving efficient update rules for the log-rank statistic is complicated by its variance calculation. In early work studying survival trees, \citet{leblanc1993survival} noted that such updates are possible for the numerator of the statistic, while \citet{leblanc1995step} later noted variance approximations that could enable constant-time updates of the denominator. 

To our knowledge, these observations have not been pursued in subsequent methodology or software, perhaps due to the perceived complexity of variance updates. In this note, we formalize these approximations into simple constant-time update rules that can be readily implemented in existing software.
This simple algorithmic refinement is implemented in the \texttt{survival\_forest} function of the \texttt{grf} \citep{GRF} package for \R \citep{Rcore}, and controlled with the \texttt{fast.logrank} argument. On a simulated example with 100 000 units observed on a 260-month time grid ($\approx$ 20 years), and 50 continuous predictors, an optimized forest runs about $3\times$ faster, while maintaining essentially identical statistical performance.

\section{Log-rank splitting}
To ease exposition, we restrict ourselves to the salient part of the random forest algorithm where the computation of the log-rank criterion plays an important role. We restrict our attention to a given node containing $n$ samples measuring the outcomes $T_i \in \{1, 2, \ldots, M\}$ that record some time point $t = 1, \ldots, M$\footnote{We can easily obtain the $M$ unique event times in any candidate node and remap them to consecutive integers $1, \ldots$ via binary search.}. We also have access to the event indicator variable $D_i \in \{0, 1\}$ indicating if the $i$-th sample is censored ($D_i=0$) or observed ($D_i=1$; this is often referred to as a ``failure time'' or ``death''). $M$ is the number of time points for which $D_i = 1$. For these $n$ samples, we consider a single candidate predictor variable $X$, which we take as given in sorted order $X_1 < X_2 \cdots < X_n$. For a given candidate split point $X_i = c$, let 
$$
L = \{i \mid X_i \leq c\}
$$
denote the set of samples in the left node and 
$$
R = \{i \mid X_i > c\}
$$
the set of samples in the right node. Let $d_{t}$ denote the number of samples with a failure event ($D_i=1$) at the $t$-th time point:
\begin{equation} \label{eq:dT}
d_t = |\{i \mid T_i = t, D_i = 1  \}|,    
\end{equation}
where we use the subscripts $L, R$ to denote the left and right node count, with:
$$
d_{t, L} + d_{t, R} = d_t.
$$
Let $Y_{t}$ denote the number ``at risk'' at the $t$-th timepoint
\begin{equation} \label{eq:atrisk}
    Y_{t} = |\{i \mid T_i \geq t  \}|,
\end{equation}
with the same left-right counterparts:
$$
Y_{t, L} + Y_{t, R} = Y_t.
$$

The log-rank criterion at the candidate split value $c$ is (as implemented and documented in \citet{rfsrc})
\begin{equation} \label{eq:logrank}
    \mathcal{L}(c) = \frac{\sum\limits_{t=1}^{M} (d_{t, L} - Y_{t, L} \alpha_t)}
    {\sqrt{ \sum\limits_{t=1}^{M} Y_{t, L}(Y_t - Y_{t, L}) \beta_t }},
\end{equation}
where we have defined the node-specific coefficients $\alpha_t$ and $\beta_t$ that stay constant for different split points $c$,
\begin{equation} \label{eq:alpha}
    \alpha_t = \frac{d_t}{Y_t},
\end{equation}
\begin{equation} \label{eq:beta}
    \beta_t = \left(\frac{Y_t - d_t}{Y_t - 1}\right) \frac{d_t}{Y_t^2}.
\end{equation}
Conceptually, this is a simple two-sample test for the null that the expected number of failures in the two nodes is the same, where under the standard sampling model $d_{t,L}$ has mean $Y_{t, L} \alpha_t$ and variance $Y_{t, L}(Y_t - Y_{t, L}) \beta_t$.

\section{Efficient log-rank splits}
A computationally demanding step in a random survival forest is scanning over all candidate split points $c$ to compute $\mathcal{L}(c)^2$ and select the split that maximizes it. A direct calculation of \eqref{eq:logrank} requires $O(M)$ operations per candidate, giving this scanning step a total runtime of $O(n M)$ for $n$ samples and $M$ distinct event times. To more closely mimic the linear-time efficiency of Breiman's original regression forests, we aim to update $\mathcal{L}(c)$ in constant time for each candidate, reducing the overall scan to $O(n)$.

\subsection{Exact numerator updates}\label{sec:numerator}
At first glance, it may appear that the numerator of $\mathcal{L}(c)$, being a sum over $t = 1, \dots, M$, must be recomputed in $O(M)$ time for each candidate. However, as noted by \citet{leblanc1993survival}, the numerator is linear in $d_{t,L}$ and $Y_{t,L}$ and can be updated in $O(1)$ time. This observation enables the linear-time scan over all split points. To see this, consider the first term of the numerator, which, using \eqref{eq:dT} and changing the order of summation, can be written as
\begin{align}
    \sum_{t=1}^{M} d_{t,L}  &= \sum_{t=1}^{M} \sum_{i=1}^{n} \one{T_i = t} \one{D_i = 1} \one{X_i \leq c} \\
    &= \sum_{i=1}^{n} \sum_{t=1}^{M} \one{T_i = t} \one{D_i = 1} \one{X_i \leq c} \\
    &= \sum_{i=1}^{n} D_i \one{X_i \leq c}.
\end{align}
Similarly, using \eqref{eq:atrisk} we get that the second term is
\begin{align}
    \sum_{t=1}^{M} Y_{t,L} \alpha_t  &= \sum_{t=1}^{M} \sum_{i=1}^{n} \one{T_i \geq t} \one{X_i \leq c} \alpha_t \\
    &= \sum_{i=1}^{n} \sum_{t=1}^{M} \one{T_i \geq t} \alpha_t \one{X_i \leq c} \\
    &= \sum_{i=1}^{n} \gamma_i \one{X_i \leq c},
\end{align}
where we have defined
\begin{equation}
    \gamma_i = \sum_{t=1}^{M} \one{T_i \geq t} \alpha_t.
\end{equation}
Together, we have that the numerator of \eqref{eq:logrank} at a given split $c$ can be written as
$$
\mathcal{L}_{\text{num}}(c) = \sum_{i=1}^{n} (D_i - \gamma_i) \one{X_i \leq c}.
$$
This form depends only on the per-sample quantities $D_i$ and $\gamma_i$, allowing a single-pass $O(n)$ update over all split points.

\subsection{Approximate denominator updates via Poissonization}
The denominator of \eqref{eq:logrank} involves hypergeometric variances, which are quadratic in $Y_{t,L}$ and cannot be updated in constant time. A simple approximation \citep[p.~105]{cox1984analysis} replaces the hypergeometric variance with
$$
\left(\frac{1}{E_1(c)} + \frac{1}{E_2(c)}\right)^{-1},
$$
where $E_1(c)$ and $E_2(c)$ are the expected number of failures in the left and right nodes. Using the $\gamma_i$ from Section~\ref{sec:numerator}, these expectations can be computed in a single pass:
\begin{align}
    E_1(c) &= \sum_{i=1}^{n} \gamma_i \one{X_i \leq c},\\
    E_2(c) &= \sum_{i=1}^{n} \gamma_i \one{X_i > c} = \sum_{i=1}^{n} \gamma_i - E_1(c)
\end{align}
The denominator can then be approximated by
$$
\mathcal{L}_{\text{den}}(c) \approx \left(\frac{1}{E_1(c)} + \frac{1}{E_2(c)}\right)^{-\frac{1}{2}},
$$
which, like the numerator, can be updated in $O(1)$ time per split point.

This approximation is well-motivated. It is well-known that the log-rank statistic is simply the score of a Cox regression. Early work recognized that Cox models could be approximated using a Poisson representation \citep{laird1981covariance, whitehead1980fitting}, but this insight appears to have been largely overlooked in modern log-rank–based survival forests. \citet{leblanc1995step} exploited this idea by Poissonizing the likelihood to derive fast approximations to the Cox score in a general step-function framework. In special cases, their score reduces to the standard log-rank statistic, with the corresponding Poissonized variance matching the formula suggested by \citet{cox1984analysis} (see \citet[Section 3]{leblanc1995step}).

\subsection{Efficient survival splitting}
Combining the exact numerator update with the approximate denominator gives what we will refer to as the LeBlanc \& Crowley splitting criterion for survival forests:
\begin{equation}\label{eq:approx_logrank}
     \mathcal{\widetilde L}(c) = \left( \sum_{i=1}^{n} (D_i - \gamma_i) \one{X_i \leq c} \right)
     \sqrt{\left(\frac{1}{E_1(c)} + \frac{1}{E_2(c)}\right)}.
\end{equation}
Algorithm~\ref{alg:logrank} implements these updates, achieving $O(n)$ evaluation of all candidate splits for a single variable\footnote{\grf has additional splitting logic for missing $X_i$-values, which essentially involves running the algorithm twice: once sending all missing values to the left, and once sending all missing values to the right (and finally, evaluating splits on missing, \citet{mayer2020doubly, twala2008good}). Practical constraints on minimum node sizes are omitted here.}.

\begin{algorithm}[ht]
    \caption{LeBlanc \& Crowley-motivated survival forest log-rank splitting rule. Scan all split points and update $\mathcal{\widetilde L}(c)$ in constant time.}
    \label{alg:logrank}
    \SetKwInOut{Input}{Input}
    \SetKwInOut{Output}{Output}
    \Input{Samples $X_i$, $i=1,\ldots ,n$, sorted in increasing order along with times $T_i$ and events $D_i$ on time index $t=1 \ldots M$.}
    \Output{Best approximate log-rank and split value.} 
    {Construct $d_t$ and $Y_t$ vectors}\\
    {Compute node-specific weights $\alpha_t \leftarrow \dfrac{d_t}{Y_t}$ }\\
    {Initialize cumulative sums $A_t \leftarrow \sum\limits_{s=1}^{M} \alpha_s \one{s \leq t}$ }\\
    {Compute node sum $\bar \gamma \leftarrow \sum\limits_{t=1}^{M} A_t$}\\
    {Initialize numerator $\mathcal{L}_{\text{num}} \leftarrow 0$}\\
    {Initialize $E_1 \leftarrow 0$}\\
    
    \For{$i=1 \ldots n$} {
        {Set $j$ to the $i$-th sample time }\\
        {Set $\gamma_i \leftarrow A_j$ }\\
        {Update numerator $\mathcal{L}_{\text{num}} \leftarrow \mathcal{L}_{\text{num}} +  D_i - \gamma_i $ }\\
        {Update variance components}\\
        {$E_1 \leftarrow E_1 + \gamma_i$}\\
        {$E_2 \leftarrow \bar \gamma - E_1$}\\
        {Compute $i$-th squared log-rank statistic with approximate variance}\\
        {$\mathcal{\widetilde L}^2_i \leftarrow \mathcal{L}_{\text{num}}^2\left(\dfrac{1}{E_1} + \dfrac{1}{E_2}\right)$ }
    }
    \Return{ $(\mathcal{\widetilde L}^2_{i^\star}, X_{i^\star})$ where $i^\star = \arg\max_i \mathcal{\widetilde L}^2_i$ }
\end{algorithm}

To highlight the computational difference between the exact and accelerated splitting rules, we benchmark the time to grow a single survival tree.
Since forest training is just an ensemble of such trees, this isolates the cost of evaluating splits and provides a direct comparison of the $O(nM)$ vs. $O(n)$ regimes.
For reference, the total forest training time scales linearly with the number of trees.
We draw survival times from a Poisson distribution with approximately $M = \{20, 130, 260, 500\}$ distinct events to mimic settings with monthly follow-up ranging from a few months to several decades (including one final setting with a nearly continuous grid, e.g., events measured in days). We set around 10\% of the units to be censored.
We then vary $p$ continuous covariates with a sample size of $n$, and measure the time it takes to grow a single tree on the full set of samples.
Table \ref{tab:timing} reports the mean runtimes, showing that $O(1)$ splitting yields a consistent speedup\footnote{Fitting deeper trees by relaxing the constraint on the number of events in each child node in \texttt{grf} can lead to slightly larger relative speedups and more noticeable absolute speedups; see Table \ref{tab:timing_alpha} in the appendix.}.

As noted in the introduction, a common heuristic is to restrict the number of event times by discretizing the time grid to a small value (e.g., $M=20$). Table \ref{tab:timing} shows that this is effective, as increasing $M$ substantially raises training time for the exact splitting rule at fixed $n$ and $p$, albeit at the cost of reduced temporal resolution. In contrast, the approximate criterion is essentially insensitive to $M$, allowing one to retain the full time resolution with little additional computational cost.
\begin{table}[ht!]
\centering
\footnotesize
\begin{tabular}{lllccc}
\toprule
$n$ & $p$ & $M$ & Runtime: Exact (s) & Runtime: Approx (s) & Speedup ($\times$) \\
\midrule
  20 000 & 25 & 20 & 0.15 & 0.12 & 1.26 \\ 
  50 000 & 25 & 20 & 0.46 & 0.36 & 1.26 \\ 
  250 000 & 25 & 20 & 3.34 & 2.77 & 1.21 \\ 
  20 000 & 50 & 20 & 0.32 & 0.25 & 1.28 \\ 
  50 000 & 50 & 20 & 1.00 & 0.81 & 1.24 \\ 
  250 000 & 50 & 20 & 6.97 & 5.70 & 1.22 \\ 
  20 000 & 100 & 20 & 0.67 & 0.52 & 1.28 \\ 
  50 000 & 100 & 20 & 2.08 & 1.71 & 1.21 \\ 
  250 000 & 100 & 20 & 14.20 & 11.62 & 1.22 \\ 
  \midrule
  20 000 & 25 & 130 & 0.34 & 0.14 & 2.50 \\ 
  50 000 & 25 & 130 & 0.94 & 0.42 & 2.26 \\ 
  250 000 & 25 & 130 & 5.89 & 3.05 & 1.93 \\ 
  20 000 & 50 & 130 & 0.68 & 0.28 & 2.46 \\ 
  50 000 & 50 & 130 & 1.96 & 0.88 & 2.23 \\ 
  250 000 & 50 & 130 & 11.90 & 6.18 & 1.93 \\ 
  20 000 & 100 & 130 & 1.44 & 0.59 & 2.44 \\ 
  50 000 & 100 & 130 & 4.05 & 1.87 & 2.17 \\ 
  250 000 & 100 & 130 & 23.89 & 12.35 & 1.93 \\ 
  \midrule
  20 000 & 25 & 260 & 0.58 & 0.15 & 3.90 \\ 
  50 000 & 25 & 260 & 1.56 & 0.45 & 3.48 \\ 
  250 000 & 25 & 260 & 8.22 & 3.20 & 2.57 \\ 
  20 000 & 50 & 260 & 1.20 & 0.30 & 3.96 \\ 
  50 000 & 50 & 260 & 3.23 & 0.97 & 3.32 \\ 
  250 000 & 50 & 260 & 17.40 & 6.50 & 2.68 \\ 
  20 000 & 100 & 260 & 2.33 & 0.63 & 3.68 \\ 
  50 000 & 100 & 260 & 6.45 & 1.95 & 3.32 \\ 
  250 000 & 100 & 260 & 35.33 & 13.24 & 2.67 \\ 
  \midrule
  20 000 & 25 & 500 & 1.06 & 0.16 & 6.49 \\ 
  50 000 & 25 & 500 & 2.75 & 0.48 & 5.74 \\ 
  250 000 & 25 & 500 & 13.68 & 3.44 & 3.97 \\ 
  20 000 & 50 & 500 & 2.22 & 0.33 & 6.64 \\ 
  50 000 & 50 & 500 & 5.58 & 1.02 & 5.47 \\ 
  250 000 & 50 & 500 & 27.59 & 6.98 & 3.95 \\ 
  20 000 & 100 & 500 & 4.32 & 0.66 & 6.51 \\ 
  50 000 & 100 & 500 & 11.82 & 2.13 & 5.55 \\ 
  250 000 & 100 & 500 & 57.38 & 14.05 & 4.08 \\ 
\bottomrule
\end{tabular}
\caption{\footnotesize Runtimes (in seconds) for growing a single tree (on all $n$ samples) using either the exact log-rank criterion \eqref{eq:logrank} or the approximate log-rank criterion \eqref{eq:approx_logrank} computed using Algorithm \ref{alg:logrank}, across increasing sample size $n$, covariate dimension 
$p$, and total number of events $M$. The runtimes are averaged over 10 repetitions.}
\label{tab:timing}
\end{table}

\FloatBarrier
\section{Empirical comparison of approximate and exact splits}
Algorithm \ref{alg:logrank} offers a clear computational advantage on large datasets: updating in \(O(1)\) time rather than \(O(M)\) can substantially reduce runtime. Equally important, however, is whether this speedup comes without loss of statistical performance.

\subsection{Comparing performance on benchmark data sets}\label{sec:simconcordance}
We consider prediction metrics that assess both discrimination and calibration. To compare discrimination, following \citet{ishwaranRSF}, we measure prediction error $PE_C$ as the complement of the $C$--index \citep{harrell1982evaluating} with the conditional cumulative hazard as outcomes (computed from the estimated Nelson--Aalen curves). This metric ranges from 0 (perfect concordance) to 1 (worst), with 0.5 corresponding to random chance. 
To compare calibration,  we measure prediction error with the integrated Brier score $PE_{IBS}$ \citep{graf1999assessment}. This metric also ranges from 0 to 1, with 0 denoting perfect calibration.

\begin{figure}[ht]
    \centering
    \begin{subfigure}[b]{0.475\textwidth}
        \centering
        \includegraphics[width=\textwidth]{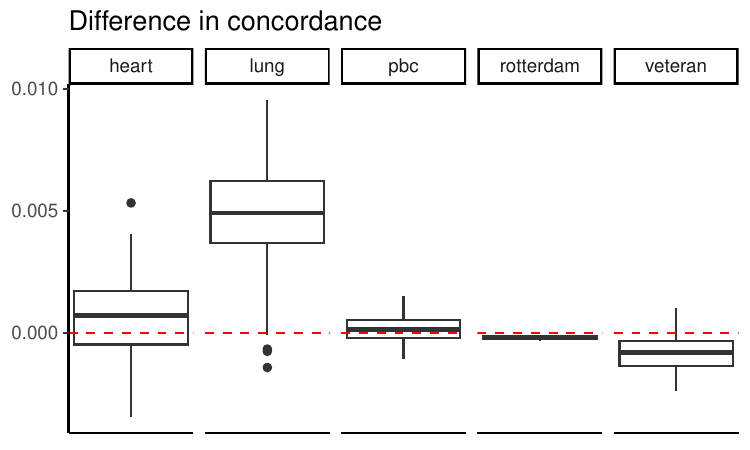}
        \caption[]
        {{}} 
        \label{fig:bench_a}
    \end{subfigure}
    \hfill
    \begin{subfigure}[b]{0.475\textwidth}  
        \centering 
        \includegraphics[width=\textwidth]{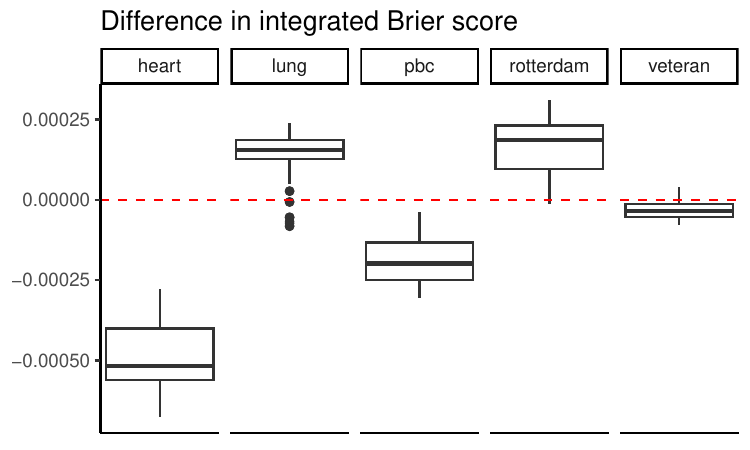}
        \caption[]
        {{}}    
        \label{fig:bench_b}
    \end{subfigure}
    \caption{\footnotesize Boxplots of difference in \ref{fig:bench_a} concordance $\Delta PE_C = PE_C(\text{exact}) - PE_C(\text{approx})$ and \ref{fig:bench_b} integrated Brier score $\Delta PE_{IBS} = PE_{IBS}(\text{exact}) - PE_{IBS}(\text{approx})$, over 250 repetitions. Typically, good survival models achieve concordance in the $0.25 - 0.35$ range and integrated Brier score in the range $0.1 - 0.3$ The prediction error difference here is orders of magnitude smaller than this baseline.}\label{fig:bench}
\end{figure}

We benchmark performance on standard survival data sets from the \texttt{survival} package \citep{survival}: Stanford heart transplant (\emph{heart}); NCCTG lung cancer (\emph{lung}); Mayo Clinic primary biliary cholangitis (\emph{pbc}); breast cancer from the Rotterdam tumor bank (\emph{rotterdam}); and the Veterans Administration lung cancer study (\emph{veteran}).
For each dataset, we compute the paired difference in prediction errors, $\Delta PE_C = PE_C(\text{exact}) - PE_C(\text{approx})$ and $\Delta PE_{IBS} = PE_{IBS}(\text{exact}) - PE_{IBS}(\text{approx})$,
by fitting random survival forests under identical settings and seeds, once with exact log-rank splits and once with approximate splits. Figure \ref{fig:bench} shows boxplots of $\Delta PE_C$ and $\Delta PE_{IBS}$ over 250 random seeds. 
Across all data sets, $\Delta PE_C$ is negligible, between $10^{-4}$ and $5\times 10^{-3}$ (an order of magnitude smaller than what's usually considered meaningful in survival model comparisons), and similarly for $\Delta PE_{IBS}$, indicating that the two algorithms are effectively indistinguishable in predictive accuracy.
Table \ref{tab:survset} in the Appendix summarizes results on a larger collection of benchmark datasets from \texttt{SurvSet} \citep{drysdale2022survset}. Across datasets and metrics, the average error difference is only around 0.00005, consistent with the results above.

\subsection{Comparing RMSE in simulated settings}
\begin{figure}[ht!]
        \centering
        \includegraphics[width=0.7\textwidth]{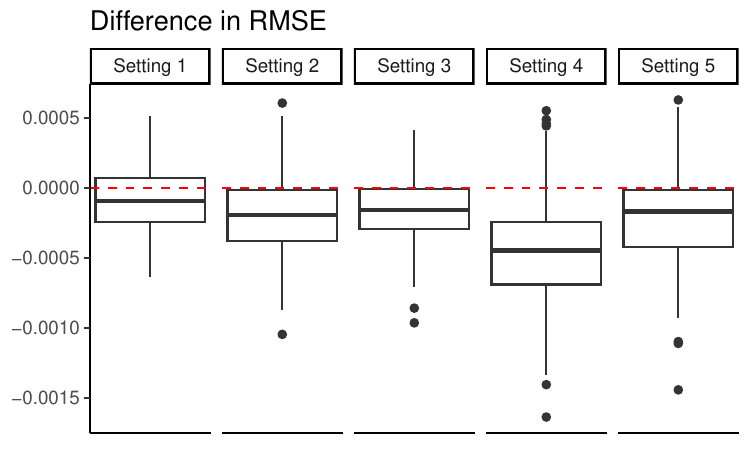}
        \caption
        {\footnotesize Boxplots of difference in prediction error $\Delta PE_{RMSE} = PE_{RMSE}(\text{exact}) - PE_{RMSE}(\text{approx})$ over 250 repetitions.
        The difference in RMSE indicates an average change in predicted survival probabilities on the order of roughly one-hundredth of a percent, effectively negligible.}
        \label{fig:rmse}
\end{figure}

We adapt the four simulation settings considered in \cite{cui2023estimating} (and add a fifth) to estimate a given time point $h$ on the conditional survival function $S(t; x)$.
We measure prediction error $PE_{RMSE} = \sqrt{\frac{1}{n} \sum_{i=1}^{n} \left( S(h; X_i) - \widehat S(h; X_i) \right)^2}$ with out-of-bag estimates.
The number of samples considered is $n$=5 000 with $p=10$ covariates. 
For details on the data generating processes, including the query time point $h$ for each setup, we refer to \citet[Section 4]{cui2023estimating}.
For each simulation setting, we compute the paired difference in prediction error, $\Delta PE_{RMSE} = PE_{RMSE}(\text{exact}) - PE_{RMSE}(\text{approx})$, as in Section \ref{sec:simconcordance}.
Figure \ref{fig:rmse} shows boxplots of $\Delta PE_{RMSE}$ over 250 simulation repetitions.  
Across all five simulation settings, the difference in RMSE between the exact and approximate criteria is on the order of $10^{-3}$, which is negligible compared to the typical RMSE values (typically measured in several percent for survival probabilities).

For completeness, we also compare runtimes on these smaller datasets in Table \ref{tab:timingbench}. The datasets in \texttt{survival} are small (except for \emph{rotterdam}, which has over 2 000 distinct event times for roughly 3 000 units) and timing differences are negligible. For the simulated datasets we see similar relative speedups as in Table \ref{tab:timing}, though at this scale the absolute time differences are unlikely to be noticeable.
\begin{table}[ht]
\centering
\footnotesize
\begin{tabular}{lccc}
\toprule
Dataset/simulation & Runtime: Exact (ms) & Runtime: Approx (ms) & Speedup ($\times$) \\
\midrule
\emph{heart} & 1.40 & 1.40 & 1.00 \\
\emph{lung} & 1.40 & 1.40 & 1.00 \\
\emph{pbc} & 1.40 & 1.40 & 1.00 \\
\emph{rotterdam} & 8.57 & 3.98 & 2.15 \\
\emph{veteran} & 1.40 & 1.38 & 1.01 \\
\midrule
\emph{Setting 1} & 31.87 & 10.47 & 3.04 \\ 
\emph{Setting 2} & 31.39 & 9.47 & 3.31 \\ 
\emph{Setting 3} & 13.24 & 10.08 & 1.31 \\ 
\emph{Setting 4} & 9.98 & 8.71 & 1.15 \\ 
\emph{Setting 5} & 31.33 & 10.00 & 3.13 \\ 
\bottomrule
\end{tabular}
\caption{\footnotesize Runtimes (in milliseconds, averaged over 100 repetitions) for growing a single tree on the datasets used in Figure \ref{fig:bench} and Figure \ref{fig:rmse}.}
\label{tab:timingbench}
\end{table}

\FloatBarrier
\section{Discussion}
Early survival tree methods were motivated by two-sample tests such as the log-rank statistic, designed for easily described prognostic subgroups in clinical settings with right-censored outcomes. In contrast, modern random forests aim to identify partitions of the covariate space that effectively separate survival distributions, rather than to optimize a single test statistic.

Approximate criteria based on first-order information have proven effective in related settings, such as gradient boosting \citep{friedman2001greedy} and generalized random forests \citep{athey2019generalized}.
Although the log-rank statistic is not explicitly gradient-based, its numerator can be viewed as a score-like quantity closely related to martingale residuals.
Consistent with this, we find that splitting on the numerator $\mathcal{L}_{\text{num}}^2$ alone yields similar empirical performance.
For the extended benchmarks in Appendix Table \ref{tab:survset}, the overall mean difference in prediction error between the exact log-rank criterion and this simplified version increases more than tenfold, but remains small in absolute magnitude.
Further improvement is obtained by standardizing $\mathcal{L}_{\text{num}}^2$ by the product of the fractions of units sent left and right, $n_L(c)n_R(c)$. This scaling more closely approximates the Poisson variance formulation, which is equivalent to replacing these fractions with the expected number of events in each node.

Prior work has also noted connections between log-rank splitting and least-squares splitting based on martingale residuals \citep{kelecs2002residual, leblanc1992relative}, further suggesting a connection to gradient-based criteria.
From a practical perspective, we prefer the approximate variance–scaled log-rank formulation in \texttt{grf}, as empirical differences are negligible and preserve compatibility with prior work using the exact criterion.

While a formal analysis is beyond the scope of this note, it would be interesting to characterize when the exact and approximate log-rank statistics induce the same rankings, and to what extent the numerator drives the split discrimination.

\section*{Acknowledgments}
We thank Ayush Kanodia for many enlightening algorithm discussions and Max Schuessler for valuable insights regarding long-term survival data in observational cancer studies. 
We are also grateful to two anonymous referees for helpful feedback.


\bibliographystyle{plainnat}
\bibliography{bibliography}

\newpage
\appendix
\section{Additional results}

\begin{table}[ht!]
\centering
\footnotesize
\begin{tabular}{lllccc}
\toprule
$n$ & $p$ & $M$ & Runtime: Exact (s) & Runtime: Approx (s) & Speedup ($\times$) \\
\midrule
  20 000 & 25 & 20 & 0.74 & 0.51 & 1.46 \\ 
  50 000 & 25 & 20 & 2.23 & 1.62 & 1.38 \\ 
  250 000 & 25 & 20 & 16.02 & 11.73 & 1.37 \\ 
  20 000 & 50 & 20 & 1.75 & 1.34 & 1.31 \\ 
  50 000 & 50 & 20 & 5.70 & 4.20 & 1.36 \\ 
  250 000 & 50 & 20 & 37.89 & 28.95 & 1.31 \\ 
  20 000 & 100 & 20 & 4.73 & 3.71 & 1.27 \\ 
  50 000 & 100 & 20 & 14.90 & 11.32 & 1.32 \\ 
  250 000 & 100 & 20 & 102.38 & 81.87 & 1.25 \\ 
  \midrule
  20 000 & 25 & 130 & 1.58 & 0.55 & 2.88 \\ 
  50 000 & 25 & 130 & 4.62 & 1.68 & 2.75 \\ 
  250 000 & 25 & 130 & 29.41 & 12.13 & 2.42 \\ 
  20 000 & 50 & 130 & 3.77 & 1.36 & 2.77 \\ 
  50 000 & 50 & 130 & 12.02 & 4.41 & 2.73 \\ 
  250 000 & 50 & 130 & 78.35 & 31.84 & 2.46 \\ 
  20 000 & 100 & 130 & 10.61 & 3.77 & 2.81 \\ 
  50 000 & 100 & 130 & 33.06 & 12.47 & 2.65 \\ 
  250 000 & 100 & 130 & 213.85 & 86.45 & 2.47 \\ 
  \midrule
  20 000 & 25 & 260 & 2.27 & 0.57 & 3.98 \\ 
  50 000 & 25 & 260 & 6.73 & 1.66 & 4.06 \\ 
  250 000 & 25 & 260 & 41.18 & 12.52 & 3.29 \\ 
  20 000 & 50 & 260 & 5.62 & 1.37 & 4.10 \\ 
  50 000 & 50 & 260 & 16.21 & 4.44 & 3.65 \\ 
  250 000 & 50 & 260 & 101.84 & 31.60 & 3.22 \\ 
  20 000 & 100 & 260 & 14.99 & 3.84 & 3.91 \\ 
  50 000 & 100 & 260 & 44.45 & 12.34 & 3.60 \\ 
  250 000 & 100 & 260 & 282.84 & 88.29 & 3.20 \\ 
  \midrule
  20 000 & 25 & 500 & 3.67 & 0.60 & 6.14 \\ 
  50 000 & 25 & 500 & 11.00 & 1.84 & 5.98 \\ 
  250 000 & 25 & 500 & 58.12 & 12.79 & 4.54 \\ 
  20 000 & 50 & 500 & 9.17 & 1.41 & 6.52 \\ 
  50 000 & 50 & 500 & 26.04 & 4.47 & 5.83 \\ 
  250 000 & 50 & 500 & 144.47 & 31.97 & 4.52 \\ 
  20 000 & 100 & 500 & 22.85 & 3.93 & 5.81 \\ 
  50 000 & 100 & 500 & 65.60 & 12.73 & 5.15 \\ 
  250 000 & 100 & 500 & 399.65 & 88.34 & 4.52 \\ 
\bottomrule
\end{tabular}
\caption{\footnotesize Timing results with deeper trees grown under less restrictive constraints on the number of events in each child node (\grf option: \texttt{alpha=0}). Runtimes (in seconds) for growing a single tree (on all $n$ samples) using either the exact log-rank criterion or the approximate log-rank criterion, across increasing sample size $n$, covariate dimension 
$p$, and total number of events $M$. The runtimes are averaged over 5 repetitions.}
\label{tab:timing_alpha}
\end{table}

\begin{table}[ht]
\centering
\footnotesize
\begin{tabular}{lllllcc}
  \toprule
& & & & & \multicolumn{2}{c}{Split constraint} \\
Dataset & n & p & M & Metric & Default & Relaxed \\
  \midrule
$\emph{hdfail}$ & 52 422 &     37 &  2 641 & $\Delta PE_C$ & 0.00000 & 0.00004 \\ 
  $\emph{prostateSurvival}$ & 14 294 &      7 &    105 & $\Delta PE_C$ & 0.00000 & 0.00076 \\ 
  $\emph{support2}$ &  9 105 &     58 &  1 041 & $\Delta PE_C$ & -0.00036 & -0.00066 \\ 
  $\emph{flchain}$ &  7 874 &     26 &  1 738 & $\Delta PE_C$ & -0.00002 & 0.00001 \\ 
  $\emph{Dialysis}$ &  6 805 &      8 &     42 & $\Delta PE_C$ & 0.00076 & 0.00181 \\ 
  $\emph{dataDIVAT1}$ &  5 943 &      6 &    810 & $\Delta PE_C$ & 0.00091 & 0.00117 \\ 
  $\emph{rhc}$ &  5 735 &     74 &    731 & $\Delta PE_C$ & 0.00039 & 0.00042 \\ 
  $\emph{Framingham}$ &  4 699 &      7 &  1 372 & $\Delta PE_C$ & -0.00051 & 0.00030 \\ 
  $\emph{dataDIVAT3}$ &  4 267 &      7 &    219 & $\Delta PE_C$ & 0.00000 & 0.00078 \\ 
  $\emph{nwtco}$ &  4 028 &      8 &    392 & $\Delta PE_C$ & 0.00026 & -0.00018 \\ 
  $\emph{smarto}$ &  3 873 &     35 &    407 & $\Delta PE_C$ & -0.00093 & 0.00173 \\ 
  $\emph{acath}$ &  3 504 &      3 &    244 & $\Delta PE_C$ & 0.00015 & -0.00014 \\ 
  $\emph{divorce}$ &  3 371 &      5 &    887 & $\Delta PE_C$ & 0.00002 & 0.00001 \\ 
  $\emph{UnempDur}$ &  3 241 &      7 &     26 & $\Delta PE_C$ & -0.00037 & -0.00105 \\ 
  $\emph{rott2}$ &  2 982 &     13 &  1 078 & $\Delta PE_C$ & -0.00006 & -0.00006 \\ 
  $\emph{Aids2}$ &  2 839 &     13 &    782 & $\Delta PE_C$ & -0.00049 & -0.00081 \\ 
  $\emph{scania}$ &  1 931 &      6 &  1 042 & $\Delta PE_C$ & -0.00038 & -0.00022 \\ 
  $\emph{TRACE}$ &  1 878 &      6 &    958 & $\Delta PE_C$ & -0.00018 & 0.00011 \\ 
  $\emph{dataDIVAT2}$ &  1 837 &      4 &    508 & $\Delta PE_C$ & 0.00026 & 0.00009 \\ 
  $\emph{actg}$ &  1 151 &     11 &     76 & $\Delta PE_C$ & -0.00064 & 0.00074 \\ 
  $\emph{LeukSurv}$ &  1 043 &      7 &    441 & $\Delta PE_C$ & 0.00106 & -0.00067 \\ 
  $\emph{rdata}$ &  1 040 &      7 &    513 & $\Delta PE_C$ & -0.00037 & -0.00025 \\ 
  $\emph{grace}$ &  1 000 &      5 &    110 & $\Delta PE_C$ & -0.00006 & -0.00031 \\ 
  \midrule
  $\emph{hdfail}$ & 52 422 &     37 &  2 641 & $\Delta PE_{IBS}$ & 0.00000 & -0.00046 \\ 
  $\emph{prostateSurvival}$ & 14 294 &      7 &    105 & $\Delta PE_{IBS}$ & 0.00000 & 0.00002 \\ 
  $\emph{support2}$ &  9 105 &     58 &  1 041 & $\Delta PE_{IBS}$ & -0.00034 & -0.00011 \\ 
  $\emph{flchain}$ &  7 874 &     26 &  1 738 & $\Delta PE_{IBS}$ & -0.00001 & 0.00003 \\ 
  $\emph{Dialysis}$ &  6 805 &      8 &     42 & $\Delta PE_{IBS}$ & 0.00003 & 0.00025 \\ 
  $\emph{dataDIVAT1}$ &  5 943 &      6 &    810 & $\Delta PE_{IBS}$ & -0.00110 & -0.00133 \\ 
  $\emph{rhc}$ &  5 735 &     74 &    731 & $\Delta PE_{IBS}$ & 0.00025 & 0.00003 \\ 
  $\emph{Framingham}$ &  4 699 &      7 &  1 372 & $\Delta PE_{IBS}$ & -0.00013 & 0.00007 \\ 
  $\emph{dataDIVAT3}$ &  4 267 &      7 &    219 & $\Delta PE_{IBS}$ & 0.00000 & 0.00009 \\ 
  $\emph{nwtco}$ &  4 028 &      8 &    392 & $\Delta PE_{IBS}$ & 0.00000 & -0.00001 \\ 
  $\emph{smarto}$ &  3 873 &     35 &    407 & $\Delta PE_{IBS}$ & 0.00008 & 0.00004 \\ 
  $\emph{acath}$ &  3 504 &      3 &    244 & $\Delta PE_{IBS}$ & -0.00009 & 0.00001 \\ 
  $\emph{divorce}$ &  3 371 &      5 &    887 & $\Delta PE_{IBS}$ & 0.00001 & -0.00002 \\ 
  $\emph{UnempDur}$ &  3 241 &      7 &     26 & $\Delta PE_{IBS}$ & -0.00006 & -0.00009 \\ 
  $\emph{rott2}$ &  2 982 &     13 &  1 078 & $\Delta PE_{IBS}$ & 0.00040 & -0.00020 \\ 
  $\emph{Aids2}$ &  2 839 &     13 &    782 & $\Delta PE_{IBS}$ & -0.00009 & 0.00008 \\ 
  $\emph{scania}$ &  1 931 &      6 &  1 042 & $\Delta PE_{IBS}$ & -0.00001 & -0.00005 \\ 
  $\emph{TRACE}$ &  1 878 &      6 &    958 & $\Delta PE_{IBS}$ & 0.00001 & 0.00004 \\ 
  $\emph{dataDIVAT2}$ &  1 837 &      4 &    508 & $\Delta PE_{IBS}$ & 0.00009 & 0.00001 \\ 
  $\emph{actg}$ &  1 151 &     11 &     76 & $\Delta PE_{IBS}$ & -0.00000 & 0.00001 \\ 
  $\emph{LeukSurv}$ &  1 043 &      7 &    441 & $\Delta PE_{IBS}$ & -0.00005 & -0.00015 \\ 
  $\emph{rdata}$ &  1 040 &      7 &    513 & $\Delta PE_{IBS}$ & -0.00013 & 0.00001 \\ 
  $\emph{grace}$ &  1 000 &      5 &    110 & $\Delta PE_{IBS}$ & -0.00008 & 0.00032 \\ 
   \bottomrule
\end{tabular}
\caption{\footnotesize Difference in prediction error (Section 4.1) using benchmark datasets from \texttt{SurvSet} \citep{drysdale2022survset}. For each dataset with at least 1 000 observations and no time-varying covariates, we fit a survival forest with identical options except for the splitting rule, and report prediction error. Some datasets have very low event rates, and the last column reports prediction error with deeper trees grown under less restrictive constraints on the number of events in each child node (\grf option: \texttt{alpha=0}).}
\label{tab:survset}
\end{table}

\end{document}